\documentclass[a4paper,conference]{IEEEtran}
\IEEEoverridecommandlockouts
\usepackage{subcaption}
\usepackage{cite}
\usepackage{amsmath,amssymb,amsfonts}
\usepackage{algorithmic}
\usepackage{graphicx}
\usepackage{textcomp}
\usepackage{balance}
\usepackage{graphicx,multirow}
\usepackage[table]{xcolor}
\usepackage[nomessages]{fp}
\usepackage{graphicx,multirow}
\usepackage{array}
\usepackage[table]{xcolor}
\usepackage{array}
\usepackage{pgfplots, pgfplotstable}
\usetikzlibrary{patterns}
\usepackage{multirow}
\usepackage{adjustbox}
\def\BibTeX{{\rm B\kern-.05em{\sc i\kern-.025em b}\kern-.08em
		T\kern-.1667em\lower.7ex\hbox{E}\kern-.125emX}}
\begin{document}
	
	\title{Sequence-to-Sequence Model with Transformer-based Attention Mechanism and Temporal Pooling for Non-Intrusive Load Monitoring}
	
	\author{\IEEEauthorblockN{Mohammad Irani Azad}
		\IEEEauthorblockA{\textit{Faculty of ECE} \\
			\textit{Qom University of Technology}\\
			Qom, Iran \\
			iraniazad.m@qut.ac.ir}
		\and
		\IEEEauthorblockN{Roozbeh Rajabi}
		\IEEEauthorblockA{\textit{Faculty of ECE} \\
			\textit{Qom University of Technology}\\
			Qom, Iran \\
			rajabi@qut.ac.ir}
		\and
		\IEEEauthorblockN{Abouzar Estebsari}
		\IEEEauthorblockA{\textit{School of the Built Environment and Architecture} \\
			\textit{London South Bank University}\\
			London, United Kingdom \\
			estebsaa@lsbu.ac.uk}
	}
	
	\IEEEoverridecommandlockouts
	
	\maketitle
	
	\IEEEpubidadjcol

	\begin{abstract}
		This paper presents a novel Sequence-to-Sequence (Seq2Seq) model based on a transformer-based attention mechanism and temporal pooling for Non-Intrusive Load Monitoring (NILM) of smart buildings. The paper aims to improve the accuracy of NILM by using a deep learning-based method. The proposed method uses a Seq2Seq model with a transformer-based attention mechanism to capture the long-term dependencies of NILM data. Additionally, temporal pooling is used to improve the model's accuracy by capturing both the steady-state and transient behavior of appliances. The paper evaluates the proposed method on a publicly available dataset and compares the results with other state-of-the-art NILM techniques. The results demonstrate that the proposed method outperforms the existing methods in terms of both accuracy and computational efficiency.
	\end{abstract}
	
	\begin{IEEEkeywords}
		NILM,  Smart Building, Deep Learning, Transformer, Attention Mechanism.
	\end{IEEEkeywords}
	
	\section{Introduction}
	Non-intrusive load monitoring (NILM) is a technique used to separate electricity consumption at the household appliance level. Smart meters only provide total energy consumption at the building level, which may not be sufficient to influence consumer behavior. The NILM process includes data collection, feature extraction, event detection, load identification, and energy separation, and machine learning techniques are used to identify appliances and extract features during steady-state and transient conditions. The NILM system can also evaluate appliance performance over a long period, helping manufacturers improve energy efficiency. Suggestions can be sent to consumers to reduce or postpone the use of portable appliances to non-peak times to save energy \cite{SMartMeterNILM}.
	
	Monitoring the energy consumption of buildings can prevent wastage and help consumers take necessary measures. Smart meters monitor overall energy consumption, providing regular feedback to consumers, which has been shown to reduce energy use by 3\%. Additionally, instantaneous energy consumption feedback at the building level can save up to 9\% \cite{EnergyPolicy13_Standard}. Consumer behavior plays a significant role in efficient energy use, and consumers are likely to change their consumption patterns based on feedback. The NILM system is an effective solution to monitor energy consumption and identify appliances that consume the most energy, providing insights to consumers and manufacturers to improve energy efficiency \cite{SustainableCities10_Review}.
		
	There are two main types of approaches in Non-Invasive Load Monitoring (NILM) - supervised and unsupervised. The supervised approach involves training models using the power consumption data of appliances. On the other hand, unsupervised methods include Factorial hidden Markov models (FHMM), hidden Markov models (HMM) \cite{PMLR12_FHMM}, and methods based on event identification and clustering. A comprehensive review of unsupervised NILM methods can be found in \cite{EEEIC15_Unsupervised,POWERCON18_OverviewNILM}.
	
	In recent years, various neural network-based methods under the supervision of NILM have been presented, thanks to the development of deep neural networks. Significant progress has been made through the use of Convolutional Neural Networks (CNN) \cite{AAAI18_Seq2Point,AAAI19_SubtaskGated}. One method, called WaveNILM \cite{ICASSP19_WaveNILM}, is based on a gated version of the dilated causal convolutional layer (DC-CNN) and examines the benefits of additional input signals while maintaining causality. Another method, proposed in \cite{EnergyBuildings22_VAE} that used a variational autoencoder, consists of two main parts: Net-IBN and VAE. Net-IBN is used to extract relevant features from raw power consumption measurements, and these features are then used in the VAE model to estimate the power consumption of each electrical device.
	
	In a different study \cite{ARxiv_COLD}, a neural network-based structure called Concurrent Loads Disaggregator (COLD) was developed to solve the NILM problem. The COLD network is based on a feedforward ReLu network with a self-attention mechanism that estimates any continuous function. The input of the spectrogram matrix network is obtained from the Short Time Fourier Transform (STFT) and is related to cumulative consumption information, and the output of the network is binary vectors that indicate the activity or inactivity of electrical devices at each time step.
	
	Another method based on transformers, called ELECTRIcity \cite{Sensors22_ELECTRIcity}, extracts features from the cumulative signal of electricity consumption. During the pre-processing stage, the model consists of a transformer-based generator and a discriminator to improve performance. The generator generates artificial signals for electrical devices using cumulative signals, and the discriminator distinguishes the artificial data produced by the generator from real data. During the training phase, the pre-trained transformer is fine-tuned in a supervised manner to predict the amount of power consumed by the electrical appliances.
	
	In a general sense, the goal of another study is to separate the consumption of electrical appliances based on the amount of consumption by the entire household. The architecture considered in this study is a combination of ResNet and Dialed Convolution network architecture. ResNet solves the gradient fading problem that occurs when the number of layers in a network increases, and Dialed Convolution is used instead of pooling layers to extract local information with less information loss \cite{ElectricPower19_deepDilated}.
	
	Overall, these methods show promising results in the field of NILM, and further research could lead to even more efficient and accurate methods for energy consumption estimation. In the paper a method for improving non-intrusive load monitoring (NILM) using various techniques such as attention mechanism, temporal pooling, residual connections, and transformers is proposed. Attention mechanism helps the model focus on relevant parts of input sequence, temporal pooling combines representations of multiple time steps into a single representation, residual connections bypass one or more layers to make gradient flow smoother during training, and transformers have been shown to achieve advanced results in NLP and time series prediction tasks. These techniques can help improve the accuracy and robustness of energy decomposition at the device level while reducing computational cost.
	
	This paper is structured as follows. In Section II, we present the proposed method, which includes the use of Transformers and Residual Connections, Temporal Pooling, and Attention Mechanism. In Section III, we describe the experimental setup and present the results obtained from applying the proposed method to several publicly available datasets. In Section IV, we provide a discussion of the results and draw conclusions regarding the effectiveness of the proposed method. We also discuss possible directions for future work in this area.
	
	\section{Proposed Method}\label{sec:review}
	The proposed method is composed of four main parts: Attention Mechanism, Temporal Pooling, Residual Connections, and Transformers. This section provides a detailed explanation of the structure of the model used.
	
	Firstly, the Attention Mechanism is a technique that allows the model to selectively focus on the most relevant parts of the input sequence at a given time step. This mechanism calculates the output element as a weighted sum of input elements, where the weights are determined dynamically by the model based on the input sequence and the current state of the model. This approach significantly enhances the performance of sequence-to-sequence models.
	
	Secondly, Temporal Pooling is a technique that extracts useful information from sequential data, such as video or speech signals. It combines representations of multiple frames or time steps into a single representation that captures the crucial features of the entire sequence. This technique addresses the challenge of processing sequences of variable length and allows the model to work with fixed-size inputs by summarizing the sequence's information in a compact representation.
	
	Thirdly, Residual Connections or coupling, is a technique that connects the output of one layer to the input of the next layer, bypassing one or more layers in between. This method makes the gradient flow more smoothly during training, avoiding the vanishing gradient problem in deep networks. It also allows the network to learn a residual mapping, which is easier to learn than the full mapping from input to output, especially when the network is very deep.
	
\begin{figure*}[htbp]
		\centerline{\includegraphics[width=13cm]{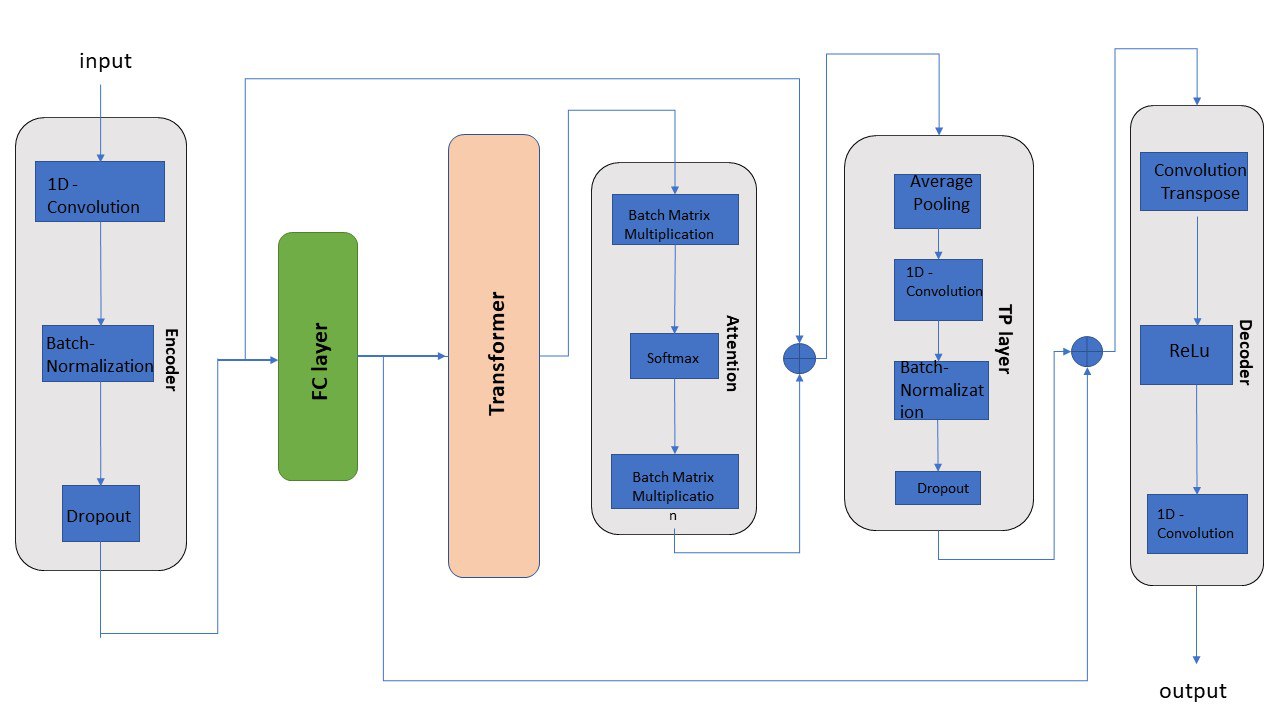}}
		\caption{Proposed sequence to sequence model using transformer-based attention mechanism and temporal pooling.}
		\label{fig:ProposedModel}
	\end{figure*}

	Finally, Transformers have been applied recently in non-intrusive load monitoring to separate household power consumption into individual appliances, achieving advanced results on a wide range of NLP and time-series prediction tasks. Overall, the proposed method presents a robust and accurate technique for device-level energy decomposition in NILM, reducing the computational cost and improving the accuracy of energy decomposition. Figure \ref{fig:ProposedModel} shows the proposed model architecture.
	
	\section{Dataset and Results}\label{sec:deepMethods}

	\subsection{Dataset}
	The UK-DALE dataset is a widely used dataset in the field of non-intrusive load monitoring (NILM) and is used extensively in this article. It consists of energy consumption data collected from five different households in England. The dataset was collected using smart meters that were installed in each household, and the data was sampled at a frequency of 1 Hz. Along with the energy consumption data, device data samples were also collected every 6 seconds, providing a detailed picture of the energy usage patterns of each household \cite{UK-DALE}.
	
	The dataset is divided into two parts: training and testing. The training dataset consists of energy consumption and device data samples from four households, while the remaining household is reserved for testing purposes. The dataset contains a total of 112 days of data for the training set and 28 days of data for the testing set. Each household in the dataset has a unique set of appliances, which provides a diverse range of energy usage patterns to train and test the NILM algorithms.

	\subsection{Experiments and Results}
	
\begin{figure}[htbp]
		\centerline{\includegraphics[width=8cm]{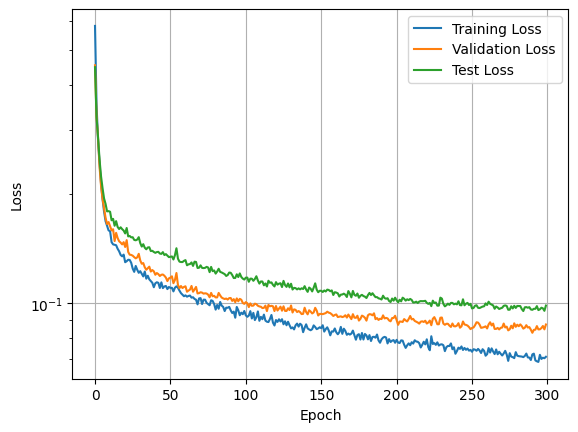}}
		\caption{Training, validation, and test losses of the proposed method for the seen case.}
		\label{fig:lossSeen}
	\end{figure}

\begin{figure}[htbp]
		\centerline{\includegraphics[width=8cm]{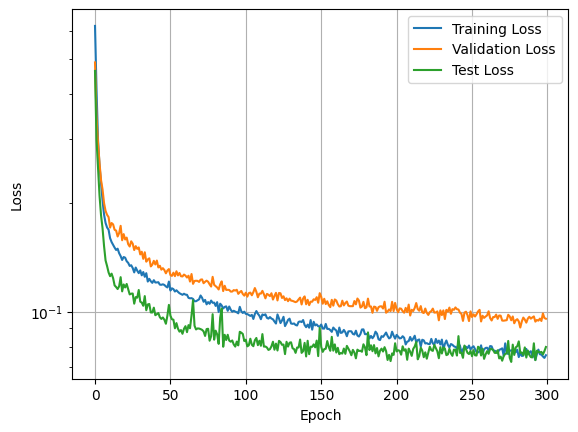}}
		\caption{Training, validation, and test losses of the proposed method for the unseen case.}
		\label{fig:lossUnseen}
	\end{figure}

	To assess the model's generalization ability and its capacity to recognize the typical characteristics of home appliances, the network was trained and tested using seen and unseen scenarios. In the seen scenario, houses used for training also included in the test set, and in the unseen scenario, houses used for training of the proposed method are not included in the test set. The model employed a multi-class classification approach for predicting the energy consumption of appliances, assuming that their consumption remains constant during operation. The network's parameters were optimized through gradient descent method using the Adam optimization algorithm with a learning rate of $10^{-4}$ and batch size of 32. The training lasted for 300 epochs. The plot of the loss per epoch during the training of the network is presented in Figures \ref{fig:lossSeen} and \ref{fig:lossUnseen} for the seen and unseen scenarios respectively. As it can be seen, in both cases the algorithm is converged and reached a satisfactory result.
	
	The results of the proposed model are summarized in Tables \ref{tab:ResultsSeen} and \ref{tab:ResultsUnseen} for seen and unseen cases respectively. It is observed that the model achieved better performance in terms of evaluation criteria such as SAE and F1 score than the existing state-of-the-art methods. These outcomes demonstrate the effectiveness of the designed model in non-intrusive load monitoring.
	
	\begin{table*}[tb]
		\centering
		\caption{Results of the proposed method for different appliances based on standard metrics in seen case.}
		\label{tab:ResultsSeen}
		\begin{tabular}{|c|c|c|c|c|c|c|c|}
			\hline
			Metrics/Appliances & F1    & Precision & Recall & Acc   & MCC   & MAE   & SAE    \\ \hline
			Fridge             & 0.885 & 0.889     & 0.876  & 0.896 & 0.791 & 13.89 & -0.019 \\ \hline
			Dish Washer        & 0.923 & 0.928     & 0.930  & 0.996 & 0.922 & 20.68 & -0.016 \\ \hline
			Washing Machine    & 0.981 & 0.976     & 0.984  & 0.997 & 0.979 & 41.87 & -0.072 \\ \hline
		\end{tabular}
	\end{table*}

	\begin{table*}[tb]
		\centering
		\caption{Results of the proposed method for different appliances based on standard metrics in unseen case.}
		\label{tab:ResultsUnseen}
		\begin{tabular}{|c|c|c|c|c|c|c|c|}
			\hline
			Metrics/Appliances & F1    & Precision & Recall & Acc   & MCC   & MAE   & SAE    \\ \hline
			Fridge             & 0.877 & 0.894     & 0.861  & 0.909 & 0.805 & 16.76 & -0.037 \\ \hline
			Dish Washer        & 0.818 & 0.801     & 0.900  & 0.991 & 0.797 & 32.38 & 0.078 \\ \hline
			Washing Machine    & 0.861 & 0.846     & 0.867  & 0.996 & 0.845 & 8.44 & 0.033 \\ \hline
		\end{tabular}
	\end{table*}

	\subsection{Comparison Study}\label{subsec:results}
	Based on the comparison of various evaluation criteria including SAE and F1 score, the results presented in Table \ref{tab:Comparison} suggest that the proposed model outperforms the previous methods. This indicates the potential of the proposed model in effectively addressing the problem of non-invasive load monitoring.

	\begin{table*}[]
		\centering
		\caption{Comparison of the proposed method with other NILM methods.}
		\label{tab:Comparison}
		\begin{tabular}{cccccccc}
			\hline
			Model                                                                      & \textbf{Metric} & DW     & FR     & KE    & MW    & WM     & \textbf{Overall} \\ \hline
			\multirow{4}{*}{WaveNILM}                                                  & acc             & -      & -      & -     & -     & -      & 94.7             \\
			& MAE             & -      & -      & -     & -     & -      & -                \\
			& SAE             & -      & -      & -     & -     & -      & -                \\
			& F1(\%)          & -      & -      & -     & -     & -      & -                \\ \hline
			\multirow{4}{*}{VAE-NILM}                                                  & acc             & -      & -      & -     & -     & -      & -                \\
			& MAE             & 23.4   & 21.6   & 22.1  & 10.8  & 6.7    & 16.9             \\
			& SAE             & -      & -      & -     & -     & -      & -                \\
			& F1(\%)          & 32.1   & 80.6   & 73.5  & 64.6  & 87.1   & 67.6             \\ \hline
			\multirow{4}{*}{COLD}                                                      & acc             & -      & -      & -     & -     & -      & -                \\
			& MAE             & -      & -      & -     & -     & -      & -                \\
			& SAE             & -      & -      & -     & -     & -      & -                \\
			& F1(\%)          & -      & -      & -     & -     & -      & 94.55            \\ \hline
			\multirow{4}{*}{ELECTRIcity}                                               & acc             & 98.4   & 84.3   & 99.9  & 99.6  & 99.4   & 96.32            \\
			& MAE             & 18.96  & 22.61  & 9.26  & 6.28  & 3.65   & 12.152           \\
			& SAE             & -      & -      & -     & -     & -      & -                \\
			& F1(\%)          & 81.8   & 81.0   & 93.9  & 27.7  & 79.7   & 72.82            \\ \hline
			\multirow{4}{*}{D-ResNet}                                                  & acc             & 98.8   & 99.6   & 99.8  & 100   & 99.6   & 99.56            \\
			& MAE             & 7.8    & 2.627  & 2.518 & 1.505 & 2.966  & 3.48             \\
			& SAE             & 0.010  & 0.020  & 0.024 & 0.162 & 0.072  & 0.0576           \\
			& F1(\%)          & 79.6   & 99.4   & 85.9  & 97.8  & 82.6   & 89.06            \\ \hline
			\multirow{4}{*}{LDwA}                                                      & acc             & -      & -      & -     & -     & -      & -                \\
			& MAE             & 6.57   & 13.24  & 5.69  & 3.79  & 7.26   & 7.31             \\
			& SAE             & 3.91   & 6.02   & 3.74  & 2.98  & 4.87   & 4.30             \\
			& F1(\%)          & 68.99  & 87.01  & 99.81 & 67.55 & 71.94  & 79.06            \\ \hline
			\multirow{4}{*}{\begin{tabular}[c]{@{}c@{}}Proposed\\ method\end{tabular}} & acc             & 99.6   & 89.6   & -     & -     & 99.7   & 96.33            \\
			& MAE             & 20.68  & 13.89  & -     & -     & 41.87  & 25.48            \\
			& SAE             & -0.016 & -0.019 & -     & -     & -0.072 & -0.035           \\
			& F1(\%)          & 92.3   & 88.5   & -     & -     & 98.1   & 92.96            \\ \hline
		\end{tabular}
	\end{table*}

	\section{Conclusion and Future Works}\label{sec:conclusion}
	In conclusion, the proposed model in this paper shows improved performance in non-invasive load monitoring compared to previous methods. The incorporation of attention mechanism, time composition, residual connections, and transformers in the model allows for a better depiction of the complex temporal patterns and dependencies of energy consumption data. The model was able to accurately identify the consumption patterns of each household appliance, even in situations where there was a significant overlap between their energy consumption patterns.
	
	The implications of accurately identifying the energy consumption patterns of personal appliances in a household are significant for energy-saving challenges, as it can help families better understand their energy consumption and identify areas where they can reduce their energy consumption. However, there is still room for future research to improve the proposed model's performance, optimize it for accuracy and computational efficiency, evaluate it on larger and more diverse data sets, and compare it with other advanced models and methods.
	
	Furthermore, incorporating additional features such as time, day, weather conditions, etc., and extending the model to multi-task learning settings or combining it with other techniques such as transfer learning and group learning can contribute to the development of more accurate and efficient energy management systems. Overall, these future research directions can contribute to the advancement of the field of non-invasive load monitoring and have important implications for energy-saving and sustainability.

	\balance
	\bibliographystyle{IEEEtran}
	\bibliography{refs}
\end{document}